\documentclass[aps,pre,twocolumn,showpacs,superscriptaddress,footinbib]{revtex4-1}
\usepackage{graphicx}

\begin{document}

\title{Quasispecies dynamics with network constraints}

\author{Valmir C. Barbosa}
\affiliation{Programa de Engenharia de Sistemas e Computa\c c\~ao, COPPE,
Universidade Federal do Rio de Janeiro,
Caixa Postal 68511, 21941-972 Rio de Janeiro - RJ, Brazil}

\author{Raul Donangelo}
\affiliation{Instituto de F\'\i sica,
Universidade Federal do Rio de Janeiro,
Caixa Postal 68528, 21941-972 Rio de Janeiro - RJ, Brazil}
\affiliation{Instituto de F\'\i sica, Facultad de Ingenier\'\i a,
Universidad de la Rep\'ublica,
Julio Herrera y Reissig 565, 11.300 Montevideo, Uruguay}

\author{Sergio R. Souza}
\affiliation{Instituto de F\'\i sica,
Universidade Federal do Rio de Janeiro,
Caixa Postal 68528, 21941-972 Rio de Janeiro - RJ, Brazil}
\affiliation{Instituto de F\'\i sica,
Universidade Federal do Rio Grande do Sul,
Caixa Postal 15051, 91501-970 Porto Alegre - RS, Brazil}

\begin{abstract}
A quasispecies is a set of interrelated genotypes that have reached a situation
of equilibrium while evolving according to the usual Darwinian principles of
selection and mutation. Quasispecies studies invariably assume that it is
possible for any genotype to mutate into any other, but recent finds indicate
that this assumption is not necessarily true. Here we revisit the traditional
quasispecies theory by adopting a network structure to constrain the occurrence
of mutations. Such structure is governed by a random-graph model, whose single
parameter (a probability $p$) controls both the graph's density and the dynamics
of mutation. We contribute two further modifications to the theory, one to
account for the fact that different loci in a genotype may be differently
susceptible to the occurrence of mutations, the other to allow for a more
plausible description of the transition from adaptation to degeneracy of the
quasispecies as $p$ is increased. We give analytical and simulation results for
the usual case of binary genotypes, assuming the fitness landscape in which a
genotype's fitness decays exponentially with its Hamming distance to the wild
type. These results support the theory's assertions regarding the adaptation of
the quasispecies to the fitness landscape and also its possible demise as a
function of $p$.
\end{abstract}


\maketitle

\section{Introduction}

The concept of a quasispecies was introduced by Eigen and Schuster
\cite{e71,es77} to describe the equilibrium state of a population of genotypes
whose members mutate frequently into one another while replicating without
recombination (i.e., asexually). At first the theory targeted the dynamics of
complex, prebiotic molecules and aimed to explain the phenomena of
self-organization and adaptability that led to the appearance of life. Today,
however, the quasispecies theory is thought to be much more widely applicable,
as to the dynamics of RNA viruses and in cancer research \cite{mlcgm10}, in
fact providing interesting insight into the dynamics of any population of
genotypes, including those that replicate with recombination and mutate
relatively infrequently \cite{nm00}.

The theory combines the evolutionary principles of selection and mutation to
describe the dynamics of a population of genotypes, and in this sense
constitutes the leading manifestation of the Darwinian principles at the
molecular level. Its central tenet is that, although each individual genotype
can be ascribed a fitness that is a function of its replicative capacity, the
actual fitness effects (ranging, e.g., from strongly deleterious to highly
adaptive \cite{o03,sme04,ek07}) are a property of the population rather than of
the genotype \cite{ss88}. As we observe the dynamics of the population relative
to the so-called fitness landscape (i.e., the fitnesses of all possible
genotypes), selection operates on the entire population and can guide it toward
the landscape's peaks. In other words, even though the process of mutation
remains essentially stochastic, the population can in fact influence it because
the fittest genotypes will replicate more and lead the population to adapt to
the fitness landscape.

In the particular case of RNA viruses, and notwithstanding some degree of
controversy over how applicable the quasispecies theory is to their dynamics
(cf., e.g., \cite{d02,hm02} and more recently \cite{h10,mlcgm10}), the array of
implications to the understanding of viral diseases is notable. For example, the
theory suggests that the fitness effects of a virus population are determined
more by how free its various genotypes are to mutate than by how capable they
are to replicate. Another implication seems to be that, paradoxically,
increasing the genotypes' error rates during replication may render the virus
less pathogenic \cite{d09,la10}.

The centerpiece of the quasispecies theory is the so-called quasispecies
equation, which for each possible genotype gives the rate at which the
genotype's relative abundance varies with time in terms of all genotypes'
abundances, their fitnesses, and the rates at which genotypes mutate into one
another. We refer the reader to \cite{be06,n06}, and references therein, for a
summary of the customary assumptions and known developments. Normally a genotype
is represented as a length-$L$ string of $0$'s and $1$'s, so the number of
genotypes in the population is $2^L$. Every genotype can mutate into every
other, so essentially there is no structure constraining the occurrence of
mutations. Moreover, in general one assumes that mutations can be modeled as
occurring independently at each of a genotype's loci with the same probability
$u$ for each locus (a notable exception here is the study in \cite{sh06}, where
loci having different mutation rates are allowed, as are mutations of two or
three adjacent loci as a group, in recognition of the plausibility of such
events \cite{arws00,df00,nc00}).

In addition to the quasispecies itself, which is characterized by the genotypes'
relative abundances at equilibrium, another important observable in the theory
is the so-called error threshold, which refers to how variations in the point
mutation rate $u$ determine the population's average fitness at equilibrium. The
customary approach to determine this threshold is to concentrate on the relative
abundance of the fittest genotype, normally called the wild (or master) type,
and study how its eventual survival depends on $u$. Invariably such studies have
assumed that no genotype can mutate into the wild type and solved the resulting,
simplified version of the quasispecies equation for the minimum value of $u$
that ensures that the wild type survives. This threshold value is a function of
the wild type's fitness and of the length $L$ \cite{be06,n06,sh06}.

Here we revisit the quasispecies theory by seeking to attenuate what we perceive
to be three main sources of biological implausibility. The first one is related
to the total lack of structure constraining the possible mutations inside the
population. Recent finds indicate, to the contrary, that for some organisms not
every combination of loci can be involved in a single mutation out of a specific
genotype \cite{vpk09}. The second one has to do with the nearly ubiquitous
assumption that genotypes are equally likely to undergo a mutation at any locus.
In this case, too, there is evidence in support of locus-dependent mutation
rates \cite{df00} even though mutations do seem to occur simultaneously at
different, not necessarily contiguous, loci \cite{wg04}.

We tackle these first two issues by adopting a susceptibility model to
differentiate one locus from another as far as the occurrence of mutations at
those loci is concerned. The susceptibility of a specific locus $\ell$ is any
positive number $s_\ell$ that gets larger as genotypes become more susceptible
to the occurrence of a mutation at locus $\ell$. Given two genotypes $i$ and
$j$ that differ at locus $\ell$ and a probability parameter $p$, we use
$p^{1/s_\ell}$ both to create a random-graph model to give structure to the
evolving population in terms of whether $i$ and $j$ can mutate into each other
and to govern the dynamics of mutation if they can. Additionally, note that by
adopting a random-graph model into the quasispecies theory we are also providing
the theory with a perspective that connects it with the decade-long effort to
understand the so-called complex networks and their applications
\cite{bs03,nbw06,bkm09}.

Our third perceived source of implausibility comes from the assumptions that
underlie the common method to determine the error threshold. Such assumptions
are too stringent (no genotype mutates into the wild type) and result in a
strict threshold separating the survival of the wild type in the quasispecies
from its catastrophic demise. Rather, as suggested by the study in \cite{wr01}
and the review in \cite{la10}, we believe it might be more plausible if the two
regimes were separated by a wider interval of the control parameter ($p$, in our
case), over which the transition could occur more smoothly. In order to avoid
the same stringent assumptions that have dominated such studies so far, we start
by assuming instead that a genotype's relative abundance in the quasispecies
depends on its fitness as a power law. The accuracy of this assumption depends
on the susceptibilities of the various loci, but in the cases we investigate it
allows the average fitness of the quasispecies to be expressed analytically and
the transition between degeneracy and survival to occur smoothly.

We proceed in the following manner, assuming that genotypes are binary (as
usual) and also that a genotype's fitness decays exponentially with its Hamming
distance to the wild type. First we introduce our model in Sec.~\ref{sec:mod},
where we rewrite the quasispecies equation for the case of network-constrained
mutations and, for two distinct susceptibility scenarios, solve it approximately
under the assumption that a genotype's relative abundance and fitness are
related by a power law. Then we give computational results in Sec.~\ref{sec:res}
and also discuss the conditions for our analytical expressions to be good
approximations to the simulation data. We conclude in Sec.~\ref{sec:concl}.

\section{Model}
\label{sec:mod}

We consider binary genotypes of length $L$, that is, length-$L$ sequences of
$0$'s and $1$'s. There are thus $n=2^L$ different genotypes, numbered
$1,2,\ldots,n$. We assume that genotype $1$ comprises only $0$'s. The fitness of
genotype $i$ reflects its replication rate and here is given by $f_i=2^{-d_i}$,
where $d_i$ is the number of $1$'s in the genotype. That is, a genotype's
fitness decays exponentially with its Hamming distance to genotype $1$ (which is
then the fittest one, with $f_1=1$, or wild type). While this choice seems
reasonable, it is by no means the only possibility and many other alternatives
might be considered. We note, however, that adopting an exponential function has
allowed many of the analytical calculations that we present in this section to
be performed.

We assume that the $n$ genotypes are the nodes of a directed graph $D$ with
self-loops at all nodes. The set of in-neighbors of node $i$ in $D$ is denoted
by $I_i$ and its set of out-neighbors by $O_i$. It holds that both $i\in I_i$
and $i\in O_i$. The existence of an edge directed from node $i$ to node
$j\neq i$ means that it is possible for genotype $i$ to mutate into genotype $j$
during replication. This happens with probability $q_{ij}$. Letting $q_{ii}$ be
the probability that genotype $i$ remains unchanged during replication leads to
$\sum_{j\in O_i}q_{ij}=1$.

Let $X_i$ denote the abundance of genotype $i$ at any given time, and
similarly let $x_i=X_i/\sum_{k=1}^nX_k$ be its relative abundance. The time
derivative of $X_i$ depends on the abundance of all genotypes in $I_i$ (i.e.,
$i$ itself and those that can mutate into $i$ during replication) in such a way
that
\begin{equation}
\dot{X}_i=\sum_{j\in I_i}f_jq_{ji}X_j.
\end{equation}
Rewriting for $x_i$ yields
\begin{equation}
\dot{x}_i=\sum_{j\in I_i}f_jq_{ji}x_j-\phi x_i,
\label{eq:qs}
\end{equation}
where $\phi=\sum_{k=1}^nf_kx_k$ is the average fitness of all $n$ genotypes.
Equation (\ref{eq:qs}) is the well-known quasispecies equation, now written for
graph $D$.

In our model, both the structure of graph $D$ and the dynamics of mutation
depend on how susceptible each of the $L$ loci in a genotype is to undergo a
mutation. For $\ell=1,2,\ldots,L$, we let $s_\ell$ be a positive number that
grows with the susceptibility that a genotype undergoes a mutation at locus
$\ell$, the same for all genotypes. Thus, an edge exists in graph $D$ directed
from genotype $i$ to genotype $j$ with probability $p_{ij}$ such that
\begin{equation}
p_{ij}=p^{\sum_{\ell=1}^Lh_\ell/s_\ell},
\label{eq:pij}
\end{equation}
where $p$ is a probability parameter and $h_\ell=1$ if and only if the two
genotypes differ at locus $\ell$ ($h_\ell=0$, otherwise). Note that this
definition of $p_{ij}$ is consistent with the mandatory existence of self-loops
at all nodes of $D$, since for $j=i$ we have $h_\ell=0$ for all $\ell$ and thus
$p_{ii}=1$. If the edge from $i$ to $j$ does exist, the probability $q_{ij}$
that $i$ mutates into $j$ (or remains unchanged, if $j=i$) is proportional to
$p_{ij}$, i.e., $q_{ij}=p_{ij}/Z_i$, where $Z_i=\sum_{k\in O_i}p_{ik}$ is a
normalizing constant for genotype $i$.

Henceforth we work on the hypothesis that, at equilibrium, $x_i$ depends on the
fitness $f_i$ as a power law for every genotype $i$. That is, we assume that
$x_i=bf_i^a$ for suitable $a>0$ when $\dot{x_i}=0$. Such functional dependency
turns up in some of the cases we study (cf.~Sec.~\ref{sec:res}) and,
furthermore, facilitates some of the analytical calculations that we carry out
in this section. It immediately follows that the equilibrium value of the
average fitness is $\phi=b\sum_{h=0}^L{L\choose h}2^{-(a+1)h}$, yielding
\begin{equation}
\phi=b\left[1+2^{-(a+1)}\right]^L.
\label{eq:phi}
\end{equation}
Moreover, from the constraint $\sum_{i=1}^nx_i=1$ we obtain
$b\sum_{h=0}^L{L\choose h}2^{-ah}=1$, whence
\begin{equation}
b=(1+2^{-a})^{-L}.
\label{eq:b}
\end{equation}

We estimate the value of $a$ by resorting to a mean-field version of
Eq.~(\ref{eq:qs}), that is, one in which the expected contribution of every
genotype $j$ to $\dot{x}_i$ (not only those in $I_i$) is taken into account and
occurs according to the expected value of the mutation probability $q_{ji}$ of
genotype $j$ into genotype $i$. By definition, mutation in this case occurs with
probability proportional to $p_{ji}$, provided graph $D$ contains an edge
directed from $j$ to $i$. The latter happens with probability $p_{ji}$ as well,
so the expected value of $q_{ji}$ is $p_{ji}^2/\sum_{k=1}^np_{jk}^2$.
Equation~(\ref{eq:qs}) then becomes
\begin{equation}
\dot{x}_i=\sum_{j=1}^n\frac{f_jp_{ji}^2x_j}{\sum_{k=1}^np_{jk}^2}-\phi x_i.
\label{eq:qsa}
\end{equation}
Our estimate of $a$ comes from considering the wild type at equilibrium, that
is, from imposing $\dot{x}_1=0$ in Eq.~(\ref{eq:qsa}) and solving the resulting
equation,
\begin{equation}
\sum_{j=1}^n\frac{p_{j1}^22^{-(a+1)d_j}}{\sum_{k=1}^np_{jk}^2}
-\left[\frac{1+2^{-(a+1)}}{1+2^{-a}}\right]^L=0.
\label{eq:a}
\end{equation}

We study two susceptibility scenarios. The first one, henceforth referred to as
the uniform case, sets $s_\ell=1$ for every locus $\ell$. In this case, it
follows that $\sum_{\ell=1}^Lh_\ell/s_\ell$ in Eq.~(\ref{eq:pij}) is the Hamming
distance between genotypes $i$ and $j$, here denoted by $H_{ij}$, and therefore
$p_{ij}=p^{H_{ij}}$. The summation on $k$ appearing in Eq.~(\ref{eq:a}) becomes
\begin{equation}
\sum_{k=1}^np_{jk}^2=\sum_{h=0}^L{L\choose h}p^{2h}=(1+p^2)^L
\end{equation}
for any $j$ and the summation on $j$, since $p_{j1}=p^{d_j}$, can be similarly
written as a sum on the possible values $h$ of the Hamming distance $d_j$ to the
wild type:
\begin{equation}
\sum_{j=1}^np_{j1}^22^{-(a+1)d_j}=\sum_{h=0}^L{L\choose h}p^{2h}2^{-(a+1)h}.
\end{equation}
This yields
\begin{equation}
\frac{1+p^22^{-(a+1)}}{1+p^2}=\frac{1+2^{-(a+1)}}{1+2^{-a}},
\end{equation}
whence
\begin{equation}
2^a=\frac{1+\sqrt{1+8p^4}}{4p^2},
\label{eq:expcase1}
\end{equation}
so in the uniform case the value of the power-law exponent $a$ does not depend
on $L$. For sufficiently small $p$, we can write $2^a\approx 1/2p^2$, which by
Eqs.~(\ref{eq:phi}) and~(\ref{eq:b}) allows the equilibrium value of $\phi$, in
the uniform case, to be approximated by
\begin{equation}
\phi=\left(\frac{1+p^2}{1+2p^2}\right)^L\approx e^{-Lp^2}
\label{eq:g1}
\end{equation}
for large $L$.

\begin{figure}[t]
\vspace{0.25in}
\includegraphics[scale=0.345]{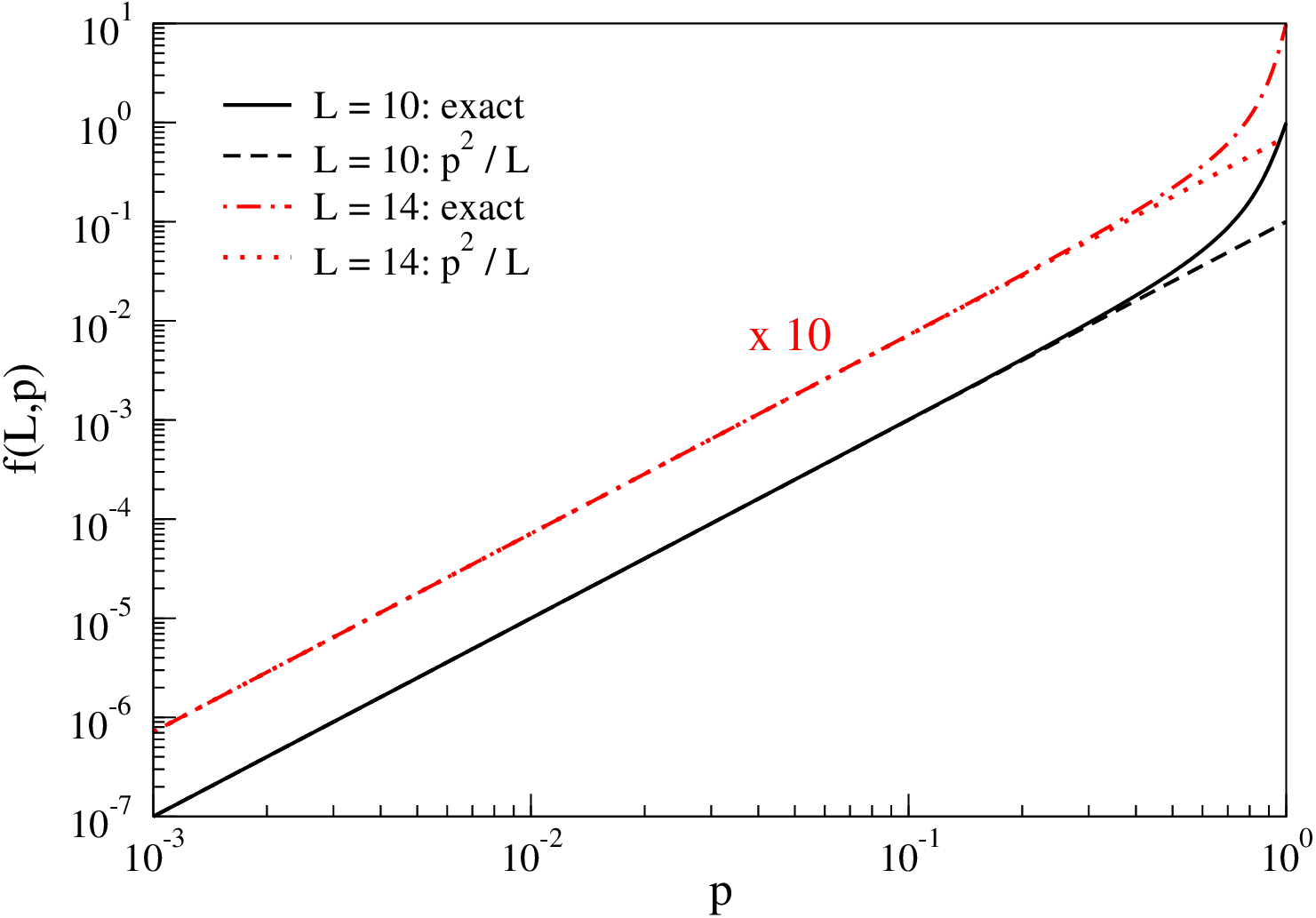}
\caption{(Color online) Approximation of $f(L,p)$ by $p^2/L$ in the
inverse-decay case.}
\label{fig:f1}
\end{figure}

In the second susceptibility scenario, which we henceforth refer to as the
inverse-decay case, we have $s_\ell=1/\ell$ for locus $\ell$. While this
specific form for the dependency of $s_\ell$ on $\ell$ is totally arbitrary and
seems to carry no special biological meaning, it has been our choice because it
is simple and has proven amenable to a certain degree of analytical
manipulation. It this case it follows that
$\sum_{\ell=1}^Lh_\ell/s_\ell=\sum_{\ell=1}^Lh_\ell\ell$ in Eq.~(\ref{eq:pij}),
which is the sum of every $\ell$ such that genotypes $i$ and $j$ differ at locus
$\ell$. Denoting this sum by $T_{ij}$ yields $p_{ij}=p^{T_{ij}}$. Now the
summation on $k$ appearing in Eq.~(\ref{eq:a}) becomes
\begin{equation}
\sum_{k=1}^np_{jk}^2=
\sum_{s=0}^{L(L+1)/2}T(L,s)p^{2s}=\prod_{\ell=1}^L(1+p^{2\ell})
\end{equation}
for any $j$, where $T(L,s)$ is the number of genotypes that differ from genotype
$j$ in loci that sum up to $s$ \footnote{Equivalently, $T(L,s)$ is the number of
partitions of $s$ into distinct parts no greater than $L$.}. The summation on
$j$, in turn, depends on first recognizing that the collective contribution to
it from all $L\choose h$ nodes $j$ whose Hamming distance to the wild type is
$d_j=h$ for fixed $h$ is proportional to
\begin{equation}
2^{-(a+1)h}\sum_{s=0}^{L(L+1)/2}T_h(L,s)p^{2s},
\label{eq:sumonj}
\end{equation}
where $T_h(L,s)$ is the number of genotypes whose $h$ $1$'s are found at loci
that sum up to $s$ \footnote{Equivalently, $T_h(L,s)$ is the number of
partitions of $s$ into exactly $h$ distinct parts no greater than $L$.}. While
the summation in this expression cannot be written in a simpler form, it can be
shown that the average value of $s$ over the $L\choose h$ genotypes involved is
$(L+1)h/2$ \footnote{Let $\protect\bar{s}$ denote the desired average. If $h=L$,
then the sum of the loci for the single possible arrangement of $1$'s is
$\protect\bar{s}=(L+1)h/2$. If $h<L$, then each of the possible arrangements is
either symmetric with respect to the genotype's center or has a symmetric
counterpart with respect to the center. Clearly, there exists a value $s$ such
that the sum of the loci equals $s$ in the former case and, moreover, the sums
of the loci in the two arrangements add up to $2s$ in the latter case. It
follows that $\protect\bar{s}=s$, so $\protect\bar{s}$ can be found by halving
the added sums of any symmetric pair. We take the pair in which the $h$ $1$'s in
each arrangement occupy outermost loci in the genotype. The corresponding sums
are $1+2+\cdots+h=(1+h)h/2$ and $(L-h+1)+(L-h+2)+\cdots+L=(2L-h+1)h/2$.
Consequently, $\protect\bar{s}=(L+1)h/2$ as well.}. We then approximate that
summation by ${L\choose h}p^{(L+1)h}$, so once again the summation on $j$ in
Eq.~(\ref{eq:a}) can be written as a sum on the possible values $h$ of the
Hamming distance $d_j$ between the wild type and genotype $j$:
\begin{equation}
\sum_{j=1}^np_{j1}^22^{-(a+1)d_j}\approx
\sum_{h=0}^L{L\choose h}p^{(L+1)h}2^{-(a+1)h}.
\end{equation}
For $f(L,p)$ such that $\prod_{\ell=1}^L(1+p^{2\ell})=[1+f(L,p)]^L$, this leads
to
\begin{equation}
\frac{1+p^{L+1}2^{-(a+1)}}{1+f(L,p)}\approx\frac{1+2^{-(a+1)}}{1+2^{-a}},
\end{equation}
and finally to
\begin{eqnarray}
2^a
&\approx&\frac{1+p^{L+1}-f(L,p)}{4f(L,p)}+\nonumber\\
&&\frac{\sqrt{[1+p^{L+1}-f(L,p)]^2+8f(L,p)p^{L+1}}}{4f(L,p)}.
\label{eq:expcase2}
\end{eqnarray}
For $p<0.2$, we have found empirically that $f(L,p)\approx p^2/L$
(Fig.~\ref{fig:f1}), whence $2^a\approx (1-p^2/L)/(2p^2/L)$ for large $L$. It
then follows from Eqs.~(\ref{eq:phi}) and~(\ref{eq:b}) that, in the
inverse-decay case, the equilibrium value of $\phi$ can be approximated by
\begin{equation}
\phi=\left(\frac{1}{1+p^2/L}\right)^L\approx e^{-p^2}.
\label{eq:g2}
\end{equation}

\section{Results}
\label{sec:res}

For fixed values of the length $L$ and the probability parameter $p$, our
results are based on generating $10^4$ independent instances of graph $D$ and
solving Eq.~(\ref{eq:qs}) numerically for each instance. This is achieved by
letting $x_i=1/n$ initially for $i=1,2,\ldots,n$ (i.e., the initial population
is uniform on all genotypes) and time-stepping the corresponding equations until
$\sum_{i=1}^n\vert\dot x_i\vert<10^{-8}$. Because this entails substantial
computational effort, we limit ourselves to $L=10$ and $L=14$ (i.e., $n=1\,024$
and $n=16\,384$ distinct genotypes, respectively).

\begin{figure}[t]
\vspace{0.25in}
\includegraphics[scale=0.345]{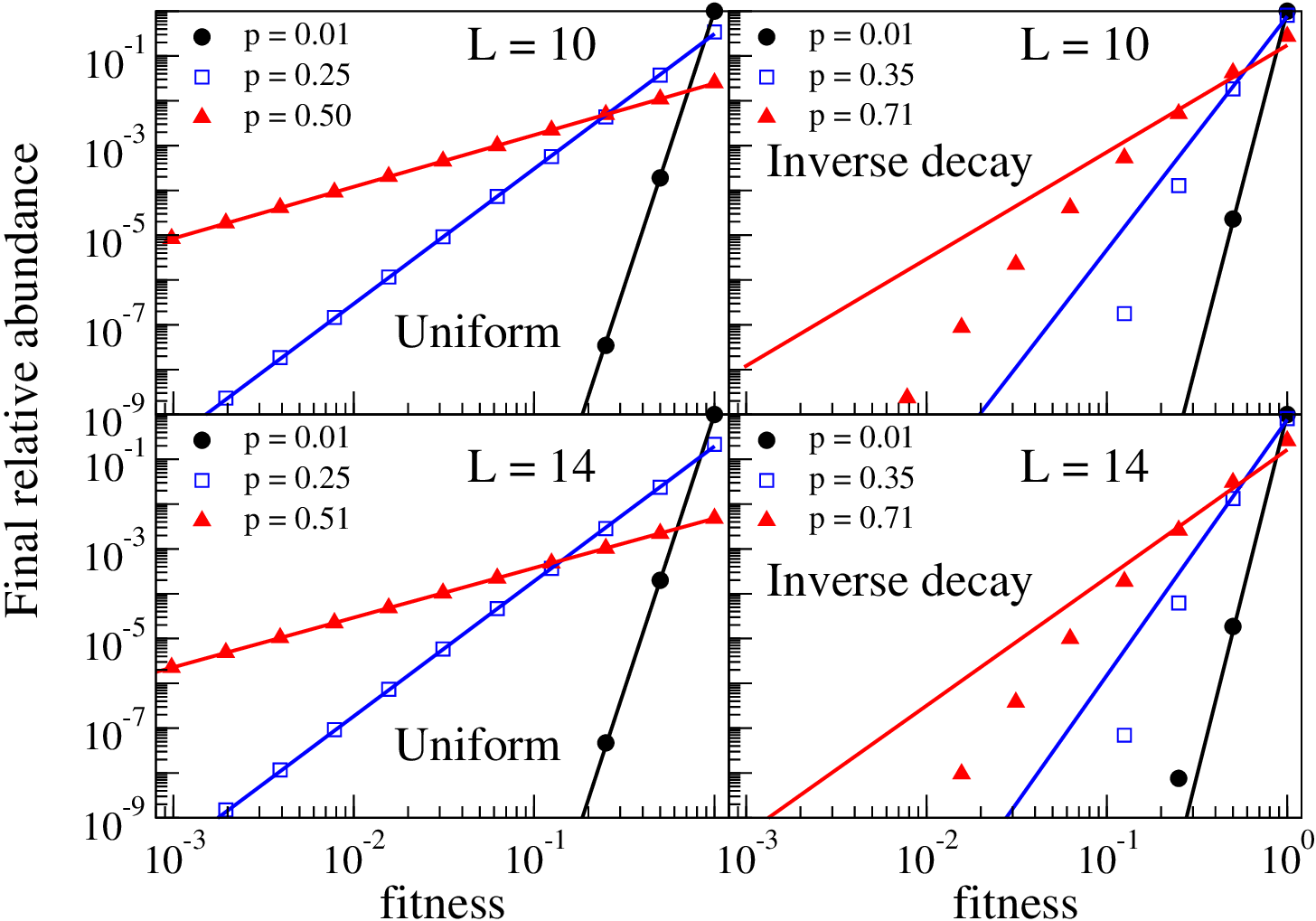}
\caption{(Color online) Relative abundances at equilibrium. For each fitness
$2^{-h}$, where $h$ is one of the possible values of the Hamming distance to the
wild type, data are averages over all $L\choose h$ genotypes that have that
fitness and $10^4$ independent instances of graph $D$. Lines refer to the power
law of exponent $a$ as given by Eq.~(\ref{eq:expcase1}) in the uniform case or
Eq.~(\ref{eq:expcase2}) in the inverse-decay case.}
\label{fig:f2}
\end{figure}

The resulting relative abundances of the quasispecies are given in
Fig.~\ref{fig:f2} as a function of the genotypes' fitnesses. By definition there
are in general several different genotypes of the same fitness, so in the figure
we give the average relative abundance of all such genotypes. In the uniform
case, these results reveal an average behavior of same-fitness genotypes that is
in excellent agreement with the power-law assumption we made. Moreover, as
indicated by Eq.~(\ref{eq:expcase1}), the power-law exponent $a$ does not depend
on $L$, being a function of $p$ exclusively. In the inverse-decay case, on the
other hand, the power-law assumption is reasonable only for the highest fitness
values. At these values, it is worth noting that the power-law exponent $a$ as
given by Eq.~(\ref{eq:expcase2}) behaves reasonably with respect to the data
despite the approximation of the summation in Eq.~(\ref{eq:sumonj}) by
${L\choose h}p^{(L+1)h}$. The reason for this is that, once these expressions
get multiplied by $2^{-(a+1)h}$ and summed up on $h$, the results are dominated
by the lowest $h$ values, hence the highest fitnesses, and these are precisely
the values at which the approximation works best [in fact, both the summation in
Eq.~(\ref{eq:sumonj}) and its approximation yield $1$ for $h=0$, since
$T_0(L,s)=1$ if $s=0$ and $T_0(L,s)=0$ otherwise].

\begin{figure}[t]
\vspace{0.25in}
\includegraphics[scale=0.345]{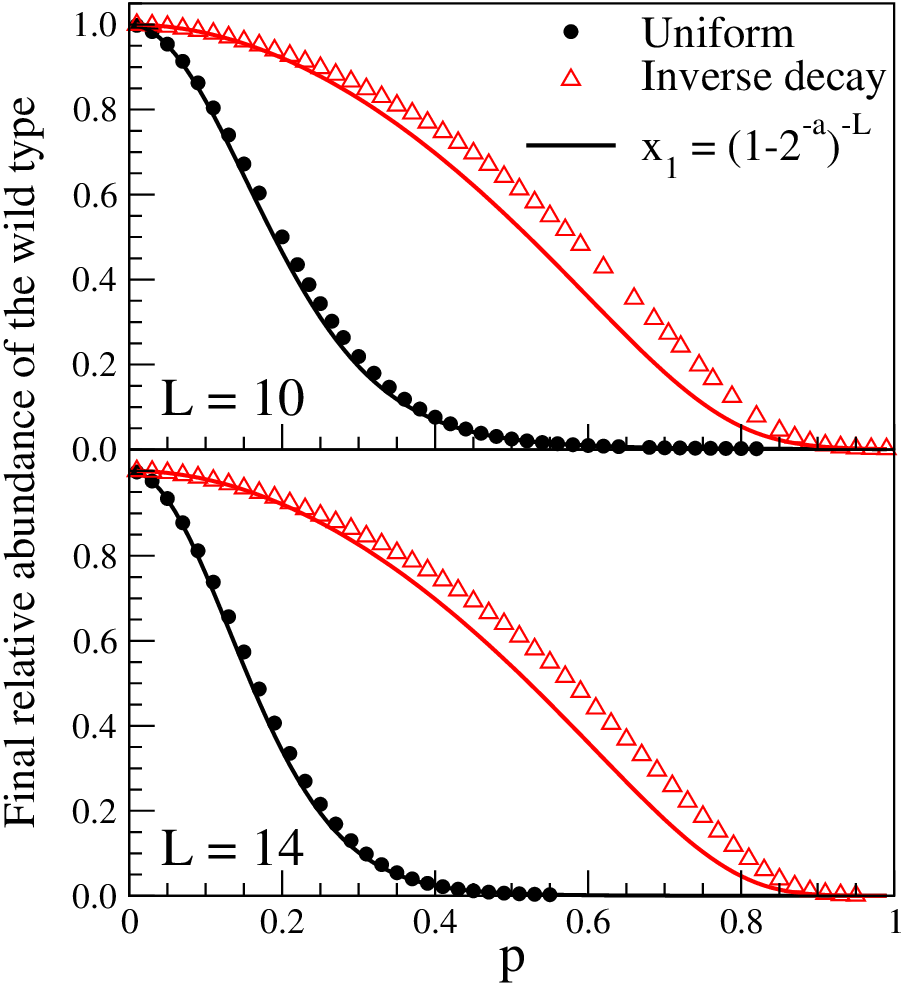}
\caption{(Color online) Relative abundance of the wild type at equilibrium. Data
are averages over $10^4$ independent instances of graph $D$. Lines refer to
$x_1=bf_1^a=b$ for $a$ as given by Eq.~(\ref{eq:expcase1}) in the uniform case
or Eq.~(\ref{eq:expcase2}) in the inverse-decay case.}
\label{fig:f3}
\end{figure}

Figure~\ref{fig:f2} also reveals how the dominance of the wild type in the
population behaves as $p$ is increased and mutations into ever more different
genotypes begin to be both allowed by the structure of $D$ and made more
frequent during the dynamics. A clearer view into this is afforded by
Fig.~\ref{fig:f3}, where we show the relative abundance of the wild type in the
quasispecies as a function of $p$. Clearly, in both the uniform and the
inverse-decay cases there exist values of $p$ beyond which the wild type gets
diluted into the population just as all other genotypes do. This happens at
higher values in the inverse-decay case, since the $1/\ell$ susceptibility for
locus $\ell$ tends to discourage mutations at this locus for all but relatively
small values of $\ell$ despite increases in $p$.

\begin{figure}[t]
\vspace{0.25in}
\includegraphics[scale=0.345]{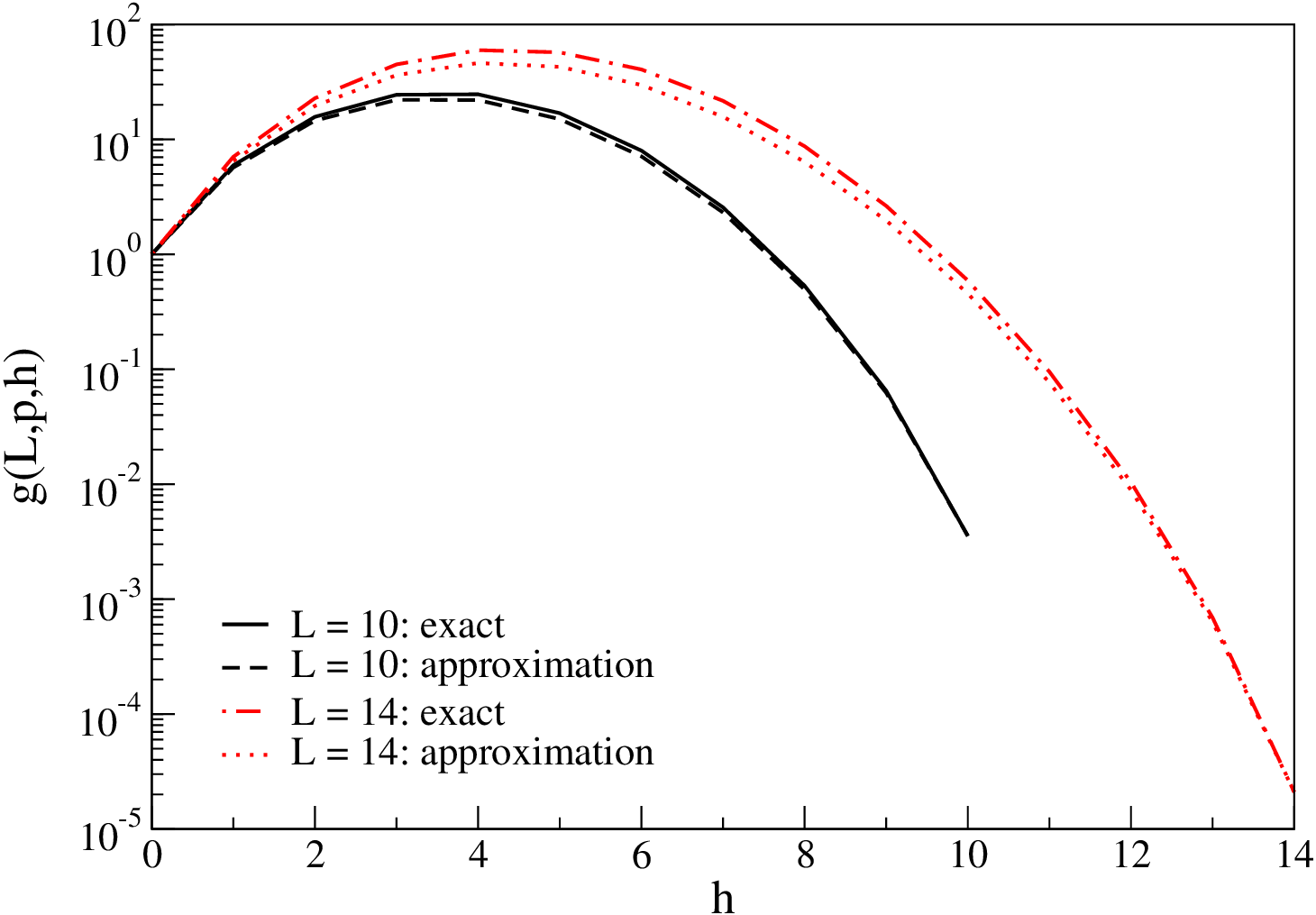}
\caption{(Color online) Comparison between the summation in
Eq.~(\ref{eq:sumonj}), here referred to as $g(L,p,h)$, and
${L\choose h}p^{(L+1)h}$ for $p=0.95$.
}
\label{fig:f4}
\end{figure}

Figure~\ref{fig:f3} also illustrates how well the power-law exponent $a$ in
Eq.~(\ref{eq:expcase1}) or~(\ref{eq:expcase2}) does when we focus on the wild
type across the entire range for $p$. While the agreement with the data is once
again very good in the uniform case, in the inverse-decay case this holds only
for roughly $p<0.2$ or $p>0.9$. As above, explaining this requires that we
revisit the approximation of the summation in Eq.~(\ref{eq:sumonj}) by
${L\choose h}p^{(L+1)h}$. Specifically, as we sum the product of either quantity
by $2^{-(a+1)h}$ on $h$, sufficiently small values of $p$ render the differences
caused by the approximation irrelevant. Similarly, for sufficiently large values
of $p$ the approximation is good across a wide range of $h$ values, as shown in
Fig.~\ref{fig:f4}.

\begin{figure}[t]
\vspace{0.25in}
\includegraphics[scale=0.345]{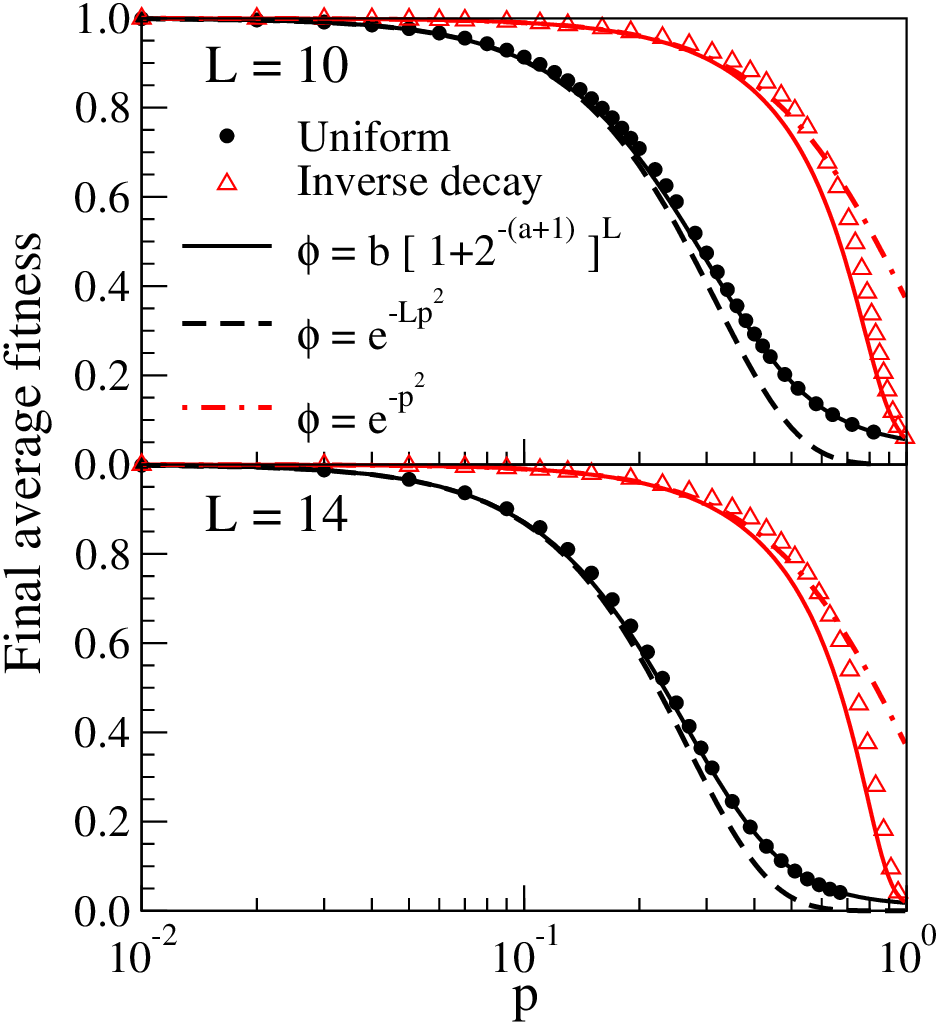}
\caption{(Color online) Average fitness at equilibrium. Data are averages over
$10^4$ independent instances of graph $D$. Lines refer to Eqs.~(\ref{eq:phi})
and~(\ref{eq:b}) with $a$ as  given by Eq.~(\ref{eq:expcase1}) in the uniform
case or Eq.~(\ref{eq:expcase2}) in the inverse-decay case, or to the Gaussian of
Eq.~(\ref{eq:g1}) in the uniform case or Eq.~(\ref{eq:g2}) in the inverse-decay
case.}
\label{fig:f5}
\end{figure}

A better glimpse into wild-type survival comes from considering the average
fitness $\phi$ of the quasispecies. This is depicted in Fig.~\ref{fig:f5}, which
clearly indicates that the transition from survival to degeneracy of the wild
type occurs gradually, within roughly one order of magnitude of the parameter
$p$ as it is increased. In the figure we also display our analytical predictions
for $\phi$ at equilibrium. These are given, through Eqs.~(\ref{eq:phi})
and~(\ref{eq:b}), as functions of the power-law exponent $a$ in
Eqs.~(\ref{eq:expcase1}) and~(\ref{eq:expcase2}). The same observations on
accuracy given above continue to apply. Figure~\ref{fig:f5} also contains the
simpler approximation of $\phi$ at equilibrium given by the Gaussians in
Eqs.~(\ref{eq:g1}) and~(\ref{eq:g2}), respectively for the uniform case and the
inverse-decay case. As expected, these approximations work very well for small
values of $p$. The one for the uniform case tends to improve as $L$ is
increased.

\section{Conclusions}
\label{sec:concl}

We have revisited the quasispecies theory and examined what we believe to be
drawbacks in its customary modeling assumptions. These are the absence of an
underlying structure separating the mutations that can occur from those that
cannot; the lack of a general framework within which a genotype's loci can be
sorted into different susceptibilities to undergo mutations; and finally, a
methodology to explain the degeneracy of the wild type, when mutations are
excessively too frequent, that implies a brusque transition from the regime in
which it survives. Our approach to tackle these issues has been, respectively,
to model the mutational interactions among genotypes as a random graph; to adopt
real-valued susceptibilities that influence both the graph's structure and the
dynamics of the population; and to postulate a specific functional dependency of
a genotype's relative abundance on its fitness at equilibrium. The resulting
model has a probability, $p$, as its single parameter. Increasing $p$ makes the
graph denser and allows more mutations as the population evolves toward the
quasispecies.

It is important to note that our model does not merely generalize the common
approach of assuming that graph $D$ has an edge directed from any genotype to
any other and that any locus in a genotype is equally susceptible to undergo a
mutation at the same point rate $u$. Even though in the two models it is
sometimes possible to write the mutation probability $q_{ij}$ of genotype $i$
into genotype $j$ as very similar products over all $L$ loci [in the customary
approach we have $q_{ij}=u^{H_{ij}}(1-u)^{L-H_{ij}}$; in our model, assuming for
example the uniform case, we have
$q_{ij}=(p/Z_i^{1/L})^{H_{ij}}(1/Z_i^{1/L})^{L-H_{ij}}$], the similarity between
them can be carried no further. In fact, setting $p=1$ in our model to ensure
that $D$ is always fully connected yields $q_{ij}=1/n=0.5^L$ regardless of $i$
or $j$, which does not conform with the usual approach unless $u=0.5$. The
bottom line is that substantial further studies are needed to determine whether
characteristic values of $p$ exist for as many organisms as possible, much as
has been done for the rate $u$ (cf., e.g., \cite{n06}).

Our results were given for the nontrivial fitness landscape in which a
genotype's fitness decays exponentially with its Hamming distance to the wild
type. They have also been based on two specific susceptibility scenarios and a
power-law relationship between a genotype's relative abundance in the
quasispecies and its fitness. While the latter is widely accurate only for one
of the susceptibility scenarios (the uniform case), overall our modeling choices
have led to useful analytical predictions of both the several genotypes'
participation in the quasispecies and the wild type's transition from survival
to degeneracy as $p$ increases.

As with other variations of the quasispecies theory, the modifications we have
introduced all corroborate the theory's central idea, viz.\ that selection and
mutation act on the entire ensemble of genotypes. They also corroborate the
crucial role of the error-related parameter ($p$, in our case) in separating two
distinct regimes, one in which the quasispecies adapts to the fitness landscape,
the other in which it becomes degenerate. It remains to be seen whether the same
will continue to hold as alternative fitness landscapes and variations of the
remaining assumptions are studied.

\begin{acknowledgments}
We acknowledge partial support from CNPq, CAPES, a FAPERJ BBP grant, FAPERGS,
the joint PRONEX initiatives of CNPq/FAPERJ under contract No.~26-111.443/2010
and CNPq/FAPERGS, and Dr.~R.~M.~Zorzenon~dos~Santos for stimulating
discussions.
\end{acknowledgments}

\bibliography{qspecies}
\bibliographystyle{apsrev4-1}

\end{document}